\documentclass{ws-procs9x6}

\newcommand{\be}{\begin{equation}}
\newcommand{\ee}{\end{equation}}
\newcommand{\ba}{\begin{eqnarray}}
\newcommand{\ea}{\end{eqnarray}}
\newcommand{\tr}{{\rm Tr\,}}
\newcommand{\ii}{{\rm i}}

\newcommand{\ex}{{\rm e}}

\newcommand{\bfx}{{\bf x}}

\newcommand{\eq}{Eq.~}

\begin{document}

\title{What mediates the longest correlation length in the QCD plasma?
}

\author{O. PHILIPSEN}
%\footnote{\uppercase{W}ork partially
%supported by grant 2-4570.5 of the \uppercase{S}wiss 
%\uppercase{N}ational \uppercase{S}cience \uppercase{F}oundation.}}

\address{Center for Theoretical Physics, MIT,\\
77 Massachusetts Ave, \\ 
Cambridge, MA 02139, USA\\ 
E-mail: philipse@lns.mit.edu}


\maketitle

\abstracts{
When thermal QCD crosses the critical temperature from below, its pressure density
rises drastically, consistent with the picture of deconfinement and the release of partons as
light degrees of freedom. On the other hand, the concept of partons
is a perturbative one, whereas interactions with the infrared modes in the plasma always
introduce non-perturbative contributions. Here I show how partonic correlators can
be defined in a gauge invariant and non-perturbative manner which applies to all energy scales.
In particular, I compute the magnetic mass for hot SU(2) gauge theory and find $m_A=0.36(2)g^2T$,
whose inverse for large $T$ is the largest correlation length in the system.
}

\section{Introduction}

\begin{figure}[th]
\centerline{\epsfxsize=6cm\epsfbox{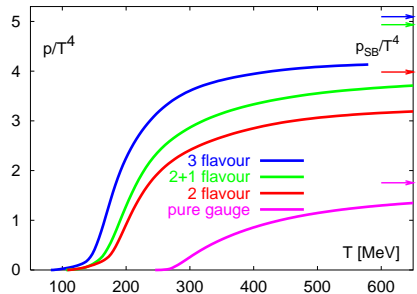}}   
\caption[]{Temperature and flavor dependence of the QCD pressure density \cite{kar}.
\label{pres}}
\end{figure}

After many years of theoretical studies and well into the period of heavy ion
collision experiments trying to establish the quark gluon plasma
of QCD, we are still far from understanding which objects constitute the
fundamental degrees of freedom in that phase, and what their properties are.
Lattice simulations of the equation of state have provided firm evidence that
the effective number of light degrees of freedom is growing rapidly 
across $T_c\sim 170$ MeV, as shown in Figure \ref{pres}.
This conclusion can be drawn because the pressure rises with temperature
for a theory with a given particle content, 
as well as with the number of light fermion flavors for all temperatures.
In the naive deconfinement picture this is explained by 
the hadronic degrees of freedom dissolving into partons, for which
one expects larger correlation lengths.

While this picture explains the qualitative features,
the flattening of the pressure short of the ideal gas value
indicates that up to $T\sim 4T_c$ interactions still play an important role. 
This in accord with the fact
that the running gauge coupling at these temperatures is still large, 
and confirmed by detailed studies of screening masses in the plasma, which 
behave non-perturbatively: contributions from the soft magnetic modes $\sim g^2T$ overpower 
those of the electric modes $\sim gT$ at the temperatures in question \cite{hlp}.
Eventually the ideal gas pressure is obtained at asymptotically
high temperatures \cite{mikko}, and the screening masses are dominated
by the perturbative contributions. But even in this regime the soft modes $\sim g^2T$ are, 
through dimensional reduction \cite{dr}, 
described by a 3d confining theory
whose perturbation theory is infrared divergent (Linde problem) \cite{ir}.
So we have evidence for the effective
degrees of freedom getting more and lighter corresponding to some kind of constituent, 
but because of the interplay between
soft and hard modes a purely perturbative parton picture is not appropriate.

To interpolate smoothly between hadronic and partonic regimes thus requires
a non-perturbative study of the dynamics of color degrees of freedom which is valid
for all scales. 
Color dynamics is encoded in Green functions of quarks and gluons, which 
in general are not gauge invariant.
In perturbation theory one fixes a gauge and studies e.g.~the field propagators directly.
While these are not physical observables,
they nevertheless carry physical information about the parton dynamics in 
their singularity structure. 
For example, the pole mass defined from the quark propagator is gauge
independent and infrared finite to every finite order in perturbation theory \cite{kro}.
A similar result holds for the gluon propagator, provided an appropriate resummation
of infrared sensitive diagrams has been performed \cite{kkr}. This has been used
to define the Debye mass $m_D\sim gT$ in analogy to QED from 
the electric gauge field propagator \cite{toni}.
Similarly, gauge invariant resummation
schemes have been designed to self-consistently compute the pole of the gluon propagator
in three dimensions \cite{bp}$^-$\cite{co}, corresponding to a ``magnetic mass'' regulating
the non-abelian thermal infrared problem. However, it has so far not been
clear whether these poles exist non-perturbatively.

On the other hand, on the lattice the study of partonic Green functions is hampered
by several problems. It is difficult to fix a gauge uniquely and avoid
the problem of Gribov copies \cite{grib}.
Moreover, most complete gauge fixings (e.g. the Landau gauge)
violate the positivity of the transfer matrix, thus obstructing
a quantum mechanical interpretation of the results. For these reasons it has 
been argued to focus on correlators of local singlet operators only. A non-perturbative
definition of the Debye mass in terms of singlet operators has been given \cite{ay}, 
identifying it as the lowest screening mass in a singlet channel odd under Euclidean time
reflection. The infrared cut-off on the magnetic scale $g^2T$ would in this picture be given
by a 3d glueball mass. However, the corresponding correlation lengths are hadronic and
not directly related to screening phenomena like e.g.~$J/\psi$-suppression,
which are caused by charged intermediate states\cite{dop}.

In this contribution I want to demonstrate that non-perturbative and gauge invariant
information is contained in an appropriately defined gluon two-point function, thus permitting
to arrive at a field theoretical definition of an associated correlation length \cite{op1,op2}.
Sections 2-4 show how to construct such an object which
can be proved to decay with eigenvalues of the Hamiltonian. Section 5 relates
the lowest eigenvalue to a level splitting between static mesons, which can be used
to compute it without recourse to any gauge fixing at all. 
Such a computation is presented for SU(2) gauge theory in 3d. In Section 6
the result is compared to those
from analytic resummation schemes, and argued to constitute the largest correlation length
in the thermal system. Section 7 presents the conclusions.

\section{A non-local gluon operator}

A gauge invariant lattice gluon correlator can be defined when a complex $N$-plet
transforming in the fundamental representation is available.
One possibility is to take the eigenfunctions
of the spatial covariant Laplacian, which is a hermitian operator with a positive spectrum,
\be \label{lev}
-\left(D_i^2[U]\right)_{\alpha\beta}f^{(n)}_\beta(x)=
%\sum_{i=1,d}\left[2f(x)-U_i(x)f^{(n)}(x+\hat{\i})-U^\dag_i(x-\hat{\i})
%f^{(n)}(x-\hat{\i})\right]=
\lambda_n f^{(n)}_\alpha(x),  \quad \lambda^n>0.
\ee
They provide a unique mapping $U\rightarrow f[U]$ except when eigenvalues are degenerate
or $|f|=0$. In simulations the probability of generating such
configurations is essentially zero.
These properties have been used previously for gauge fixing
without Gribov copies and to construct blockspins for the derivation of
effective theories \cite{lap}. 
The lowest eigenvectors are used to construct the matrix
$\Omega(x)\equiv\frac{1}{|f(x)|}\left(
f^{(1)}(x),f^{(2)}(x)\right)
$, which transforms as $\Omega^g(x)=g(x)\Omega(x)h^\dag (t)$.
Composite link and gluon fields are defined by
\ba \label{cl}
V_\mu(x)&=&\Omega^\dag(x)U_\mu(x)\Omega(x+\hat{\mu}),\\
A_\mu(x)&=&\frac{\ii}{2g}\left[V_\mu(x)-V^\dag_\mu(x)
-\frac{1}{N}\tr\left(V_\mu(x)-V^\dag_\mu(x)\right)\right],
\ea
both transforming as
$O_i^g(x)=h(t)O_i(x)h^\dag(t)$, whereas $V_0^g(x)=h(t)V_0(x)h^\dag(t+1)$.
The gauge field at a given time now has only a global gauge freedom left.
To cancel this in the correlator,
the zero momentum projected time links
$\tilde{V_0}(t)=\sum_{\bfx}V_0(\bfx,t)/|\sum_{\bfx}V_0(\bfx,t)|$
are multiplied to ``strings'' $\tilde{V}_0(t_1,t_2)$ connecting two timeslices.
These ingredients can be combined to the gauge invariant operator
\be \label{ofinal}
O[U]=\tr \left[A_i(\bfx,0)\tilde{V}_0(0,t)
A_i(\bfx,t)\tilde{V}_0^\dag(0,t)\right],
\ee
which in the particular gauge $V_0(t)=1$ reduces to the gluon propagator.
Hence we have a gauge invariant non-local observable, which is identical to the
gluon propagator in a particular gauge.

Using the transfer matrix formalism \cite{cre}, this observable can be converted to a trace
over quantum mechanical states, falling off exponentially as 
\be \label{finalen}
\langle O[U]\rangle = \sum_{n}
|\langle 0|\hat{A}_{\alpha\beta}(\bfx)
|n^{L0}\rangle |^2 \ex^{ -(E_n-E_0)t}\;,
\ee
provided that there is a finite energy gap $(E_1-E_0)$.
The eigenvalues $E_n$ and eigenvectors $|n^{L0}\rangle$ are those of the
Hamiltonian in Laplacian temporal gauge, $V_0(x)=1$.
It has been proved \cite{op1} that this Hamiltonian has the same
spectrum as the Kogut-Susskind Hamiltonian $\hat{H}_0$ \cite{ks}, which is obtained by quantizing
the theory in temporal gauge. Note that the zero momentum projection of the $V_0$ switches
off the divergent self-energy of the sources, so the energies are finite.

\section{Discussion and interpretation}

One may now ask to what extent these results depend on the 
particular choice of $\Omega[U]$, which is not unique.
It is crucial that $\Omega$ depends only on spatial links
to preserve the transfer matrix.
Clearly, any $\Omega\in SU(N)$ local in time and
transforming in the same way 
permits construction of the gauge invariant observable \eq (\ref{ofinal}). From the
spectral representation it follows that all such observables fall off with the same
spectrum, while $\Omega$ only enters the matrix elements representing
the overlap of the operator with the eigenstates.

The construction of the composite link variable \eq (\ref{cl})
may also be viewed as fixing Laplacian gauge on each timeslice.
Our result then implies
that all gauges satisfying the mentioned constraints (e.g.~the standard lattice
Coulomb gauge) will produce the same exponential decay.
Fourier transforming \eq (\ref{finalen}) to momentum space, we obtain the
non-perturbative analogue to the situation in perturbation theory: 
the energies appear as
gauge invariant poles, while the matrix elements correspond to gauge dependent 
residues. 

Note that this does {\it not} imply an asymptotically free colored state.
The temporal links in the operator
indicate the presence of static charges, as e.g.~in a Wilson loop.
In a Hamiltonian formulation \cite{ks,cre} the Kogut-Susskind Hamiltonian $\hat{H}_0$ 
acts on a Hilbert space $H_0$ of all complex,
square integrable wave functions.
Gauge invariant wave functions, $\psi[U^g]=\psi[U]$,
form a subspace $H\subset H_0$, on which the gauge invariant projected
Hamiltonian $\hat{H}$ acts.
Generally, the spectrum of $\hat{H}$ consists of the physical
particle states of the theory, which couple to local gauge invariant operators.
In addition to these states, $\hat{H}_0$ contains also the spectrum of
gauge field excitations in the presence of static sources, 
such as the static potential, gluelumps etc. These are gauge invariant energy eigenvalues
to non-trivially transforming pure gauge wave functions. Full gauge invariance is restored
once the source fields are made explicit. 
Our correlator thus probes a gauge field
excitation with the quantum numbers of the gluon in the presence of sources, 
which do not contribute to the field energy.

\section{A numerical check}

\begin{figure}[th]
%\vspace*{-0.1cm}
\centerline{\epsfxsize=5cm\hspace*{0cm}\epsfbox{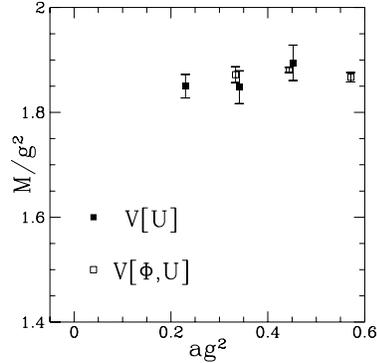}}
\caption[]{{
The W-boson mass in a 3d Higgs phase, computed from the standard operator
$V[\phi,U]$ \cite{us} and the non-local $V[U]$, \eq (\ref{cl}) \cite{op2}. The gauge coupling is 3d.}}
\label{comp}
\end{figure}

It is expedient to first test this new operator in a 3d SU(2) Higgs model,
which in its broken phase has physical states with the quantum numbers of
the gluon, the W-bosons, with detailed results available \cite{us}.
Figure \ref{comp} compares the W-mass measured in these works by the standard
operator $V[U,\phi]={\rm Im}\left(\phi^\dag(x)U(x)\phi(x+\hat{\mu})\right)$, with
results obtained from the composite links $V[U]$. Full agreement is observed
for different lattice spacings, which also confirms that the energy levels \eq (\ref{finalen})
do have a continuum limit. 
On the other hand, when simulated in the pure gauge
theory, they grow linearly with the box size \cite{op2}.
This effect stems from the non-locality of the functional $\Omega[U]$, which
depends on all link variables in a t-slice. On a periodic torus and in a
confining regime, it will thus
project predominantly on torelonic states and be blind to local quantities.

\section{The gluon propagator and static mesons}

It is then necessary to construct an alternative operator coupling to the
states we are interested in. This is achieved by introducing explicit scalar fields for
the fundamental static sources.
In continuum notation, adding
\be
S_{\phi}[U,\phi]=\sum_x\left\{-|D_\mu\phi(x)|^2 + m_0^2|\phi(x)|^2\right\}
\ee
to the pure gauge theory results in QCD with scalar quarks.
In the limit $m_0\rightarrow \infty$ the scalars become static sources propagating
in time only,
their propagator being known exactly to consist of
temporal Wilson lines $U_0(t_1,t_2)$.
Scalar and vector mesons are described by
$S(x)=\phi^\dag(x)\phi(x),V(x)={\rm Im}(\phi^\dag(x) D_i(x)\phi(x))$, respectively.
In the static limit the scalar fields do not contribute to angular
momentum, nor can they be excited into higher quantum states since they are quenched.
Consequently the mass difference $m_A\equiv \lim_{m_0\rightarrow \infty}(M_V-M_S)$ 
is a pure gauge quantity,
characterizing an excitation with the quantum numbers of the gluon.
\begin{figure}[th]
\centerline{\epsfxsize=6cm\epsfbox{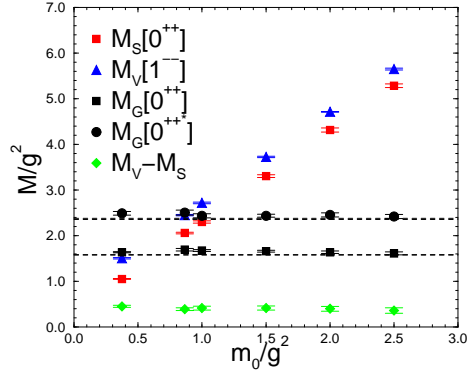}}   
\caption[]{{The lowest states in 3d scalar QCD \cite{op2}. $M_G$ denotes scalar glueballs, $M_{S,V}$
scalar and vector mesons. The gauge coupling is 3d.}}
\label{mhiggs}
\end{figure}
Moreover, integrating the scalars out analytically,
one obtains for the ratio of correlators
\be
\frac{\langle V(\bfx,0)V(\bfx,t)\rangle_c}{\langle S(\bfx,0)S(\bfx,t)\rangle_c} \sim
\frac{\int DA \;\tr\left(D_i(\bfx,0)U_0(0,t)D_i(\bfx,t)U^\dag_0(0,t)\right)
\ex^{-S_{YM}} }
{\int DA \; \tr\left(U_0(0,t)
U^\dag_0(0,t)\right)\ex^{-S_{YM}}}.
\label{ratio}
\ee
In temporal gauge this reduces to the gluon propagator again.
In other words, the mass difference $M_V-M_S$ in the static limit 
should be equivalent to the pole mass of the gluon.
(Similarly, a correlation length for thermal electric gluons was recently extracted from
a ratio of singlet correlators \cite{gg}.)

Numerical results for 2+1 dimensional SU(2) are shown in Figure \ref{mhiggs}.
With increasing scalar mass, the measured glueball states $M_G$ attain their pure
gauge values indicated by the dashed lines. The scalar bound states
move out of the spectrum, with $M_V-M_S$ approximately constant. At the largest
scalar mass one finds $m_A=0.37(6)g_3^2$, or $M_G/m_A\approx 4.2$.
%In a similar approach, an electric gluon correlation length was recently extracted from
%a ratio of singlet correlators \cite{gg}.

Rather than approaching the static limit numerically, 
one may also compute it directly,
beginning from the right hand side of \eq (\ref{ratio}). The numerator is a continuum version of
the lattice operator \eq (\ref{ofinal}), the difference being that the time-like strings are not
zero momentum projected. In discretizing the covariant derivative, the non-local functionals
should be avoided. This is achieved by observing that the exponential decay of a correlator
is entirely determined by the Hamiltonian and the quantum numbers of the operators used.
Instead of discretizing the covariant derivative, one can then equally well employ a local
higher dimension operator sharing the same quantum numbers, such as $(D_jF_{ij}(x))^a$.
The discretized version of the latter is simply the adjoint part of a 
linear combination of plaquettes, and hence a local operator.
The denominator, on the other hand, consists of a Wilson line running back and forth, 
i.e.~an adjoint Wilson line, which has the same exponential decay as a field strength
correlator \cite{mic}. The discretized version of \eq (\ref{ratio}) is then simply given
by a ratio of gluelump correlators with the appropriate quantum numbers, 
\be
\frac{\left\langle \tr\left(D_iF_{ij}(\bfx,0)U_0(0,t)
D_iF_{ij}(\bfx,t)U_0^\dag(0,t)\right)\right\rangle}
{\left\langle \tr\left(F_{ij}(\bfx,0)U_0(0,t)
F_{ij}(\bfx,t)U_0^\dag(0,t)\right)\right \rangle},
\ee
and its asymptotic exponential decay is the mass difference of the corresponding gluelump 
masses, in which the divergent self-energies of the Wilson lines cancel. (Note that, had we
coupled the gluon to one adjoint source field instead of two fundamental, we would have obtained
the same static limit.)
The calculation performed in \cite{op2}
and extrapolated to the continuum gives as the final result
\be
m_A=0.360(19)g_3^2.
\ee
Note the agreement with the result obtained by using heavy scalar fields.

\section{The magnetic mass}

In the framework of the 4d thermal field theory, the 3d gluon mass appears at asymptotic
temperatures as the magnetic mass
for hot SU(2) gauge theory, which then is $m_{A}=0.36(2)g^2T$. 
It is now interesting to compare with gauge invariant resummation
schemes of perturbation theory and a Hamiltonian strong coupling analysis,
all done in the 3d gauge theory, which have been used in the past to compute the pole of
the propagator. 
As the Table shows, the leading order results of all these calculations get within
 20\% of the right answer, and the two-loop calculation \cite{eb} in one of the schemes \cite{bp}
even suggests reasonable convergence. 
\begin{table}[ht]
\begin{center}
%\tbl{Comparison of magnetic mass values from
%gap equations and an Hamiltonian analysis.\vspace*{1pt}}
\begin{tabular}{|c|cc|cc|}
\hline
               & Ref. && $m_A/g_3^2$& \\ \hline
1-loop gap eq. & \cite{an} &   & 0.38 & \\
               & \cite{bp,jp}& & 0.28 &\\
               & \cite{co}  & & 0.25 &\\
2-loop gap eq. for \cite{bp}& \cite{eb}   & & 0.34 &\\ 
Hamiltonian strong coupl.  & \cite{nair} & & 0.32&\\\hline
\end{tabular} \label{mmag}
\end{center}
\end{table}
\begin{figure}[th]
\centerline{\epsfxsize=7cm\epsfbox{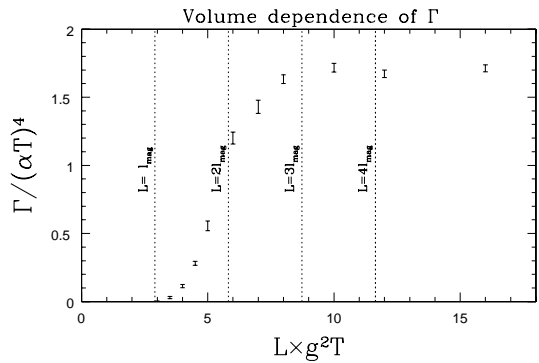}}
\caption[]{Volume dependence of the sphaleron rate, which picks up at $2l_{mag}$ (the 2 is
for the periodic boundary conditions) \cite{guy}.}
\label{sphal}
\end{figure}
For non-perturbative evidence for the role of this quantity in non-abelian plasmas,
recall Guy Moore's contribution to SEWM 2000 \cite{guy},
in which he showed the finite size scaling behavior of the sphaleron rate in 
the hot SU(2) pure gauge theory. This is obtained by simulations of the effective theory
for the soft modes $\sim g^2T$ and shown in Figure \ref{sphal}. 
A quantity is typically afflicted by finite size effets as long as the correlation length
corresponding to the relevant modes does not fit into the simulated box, $L \leq \xi$,
and begins to approach its infinite volume limit once $L>\xi$.
The sphaleron rate displays a rather clear signal for this, and the correlation length one 
estimates from the plot is fully compatible with $m_A^{-1}$, while being completely at odds
with $M_G^{-1}$.
%\footnote{It is also compatible with the spatial, or three-dimensional, 
%string tension
%$\sqrt{\sigma}^{-1}$, but $\sigma$ is the coefficient of an area and not a distance, and without
%further specification
%not an inverse correlation length.}. This is not surprising because topology is carried precisely
%by pure gauge modes of the gauge field, which cancel out in any singlet operator.  

\section{Conclusions}

By constructing a non-local lattice operator whose correlation function is amenable to the
transfer matrix formalism, I have shown that a non-perturbative, gauge invariant mass scale
is associated with the gluon field, representing the smallest eigenvalue of
the Kogut-Susskind Hamiltonian in the presence of external charges. In momentum space 
it corresponds to a pole in the propagator in all unique gauges that are local in time.
The eigenvalue can be related to some particular level splittings of static mesons.
I have calculated this quantity for the 2+1 dimensional SU(2) pure gauge theory,
and found it to be roughly a quarter of the glueball mass.
This implies a magnetic mass for the hot SU(2) gauge theory, whose
inverse constitutes the largest correlation length of the system at asymptotic
temperatures. 
Of course, 
close to $T_c$ the increase of $g^2(T)$ may lead to level crossings
with electric modes and change this picture, and so will the addition of light fermion flavors.
But non-perturbatively dressed partons are theoretically accessible and,
as the sphaleron example shows, might be the relevant degrees of freedom for certain aspects 
of plasma dynamics.

\end{document}